\documentclass[12pt,a4paper]{article}   

\newcommand{\beq}{\begin{equation}}
\newcommand{\eeq}{\end{equation}}

\def\beqa{\begin{eqnarray}}
\def\enqa{\end{eqnarray}}
\def\beq{\begin{equation}}
\def\enq{\end{equation}}
\def\nonum{\nonumber}

\begin{document}
\setlength{\baselineskip}{7mm}
\hspace*{0mm}

\par
\begin{center}
  {\large\bf Equivalence of TBA and QTM }
\end{center}
\par
\begin{center}
{\bf  Minoru Takahashi, Masahiro Shiroishi and Andreas Kl\"umper$^{a}$}\\
{\it Institute for Solid State Physics, University of Tokyo, \\
Kashiwanoha 5-1-5, Kashiwa, Chiba, 277-8581 Japan,\\
$^a$Universit\"at Dortmund, Fachbereich Physik, \\
Otto-Hahn-Str. 4, D-44221 Dortmund, Germany

 }    

\vspace*{1cm}
Abstract\\
\end{center}

\par
\setlength{\baselineskip}{6.5mm}
\setlength{\parindent}{18pt}
\noindent
The traditional thermodynamic Bethe ansatz (TBA) equations for 
the XXZ model at $|\Delta|\ge 1$ are
derived within the quantum transfer matrix (QTM) method. 
This provides further evidence of the equivalence of both methods.
Most importantly, we derive an integral equation for the free energy 
formulated for just one unknown function. This integral equation is
different in physical and mathematical aspects from the established
ones. The single integral equation is analytically 
continued to the regime $|\Delta|<1$.\\

\section{Introduction}
The thermodynamics of one-dimensional solvable models is generally 
determined by the solution to a 
set of socalled thermodynamic Bethe ansatz (TBA) equations \cite{tak99}.
Some lattice spin models such as the
XXZ chain, XYZ chain have been treated also by the quantum transfer matrix 
(QTM) method \cite{koma89, tak91a, tak91b, AK}, see also chapters 17 and 18 
of \cite{tak99}.
Correlated electron systems such as t-J model and Hubbard model, have also been treated by TBA and QTM methods
\cite{schlot87,jutt97,jutt98,tak72,koma90,jutt98b}. 

The equations obtained by the QTM approach are quite different from
those of TBA. However, the numerical results of the two methods for
the free energies are the same. Mathematically the non-linear integral
equations of \cite{tak99} and \cite{AK} share similarities insofar as they
can be interpreted as equations for dressed energies of elementary particles
of magnon and spinon type, respectively.

Recently, from the stand point of TBA one of the authors (MT)
\cite{MT} derived in the case of the XXZ chain a simple integral equation for
just one unknown function. This integral equation is completely
different in structure from those mentioned above.  Here we aim at a
derivation of this equation in the Quantum Transfer Matrix approach
providing a more explicit as well as unified understanding of the
structures and involved functions.
 
To be definite, we first consider the region ${\Delta \ge 1}$,
\begin{eqnarray}
{\cal H} &=& -J \sum_{i=1}^{N} \left\{ S_i^x S_{i+1}^x + S_i^y S_{i+1}^y + 
\Delta (S_i^z S_{i+1}^z - \frac{1}{4} ) \right\} - 2 h \sum_{i = 1}^{N} S_i^z.
\end{eqnarray}
The thermodynamic Bethe ansatz equations for this model at temperature $T$ are
called Gaudin-Takahashi equations, \cite{tak71a,gaudin71} 
\beqa
&&\ln\eta_1(x)={2\pi J\sinh\phi \over T\phi}{\bf s}(x)+
{\bf s}*\ln(1+\eta_2(x)),\nonum\\
&&\ln\eta_j(x)={\bf s}*\ln(1+\eta_{j-1}(x))(1+\eta_{j+1}(x)),
\quad j=2,3,...,\nonum\\
&&\lim_{l\to \infty}{\ln\eta_l\over l}={2h\over T}.\label{eq:GTeq}
\enqa
Here we put
\beq
\Delta=\cosh\phi,\, Q\equiv \pi/\phi,\,
{\bf s}(x)={1\over 4}\sum_{n=-\infty}^\infty 
{\rm sech}\Bigl({\pi (x-2nQ)\over 2}\Bigr) ,\,
{\bf s}*f(x)\equiv \int^Q_{-Q}{\bf s}(x-y)f(y){\rm d}y.
\enq
The free energy per site is 
\beq
f={2\pi J\sinh \phi\over \phi}\int^Q_{-Q}{\bf a}_1(x){\bf s}(x){\rm d}x
-T\int^Q_{-Q}{\bf  s}(x)\ln(1+\eta_1(x)){\rm d}x,\,
{\bf a}_1(x)\equiv{\phi\sinh\phi/(2\pi)\over \cosh \phi -\cos(\phi x)}.
\label{eq:fen1}
\enq From this equation MT \cite{MT} derived 
\beqa
&&u(x)=2\cosh({h\over T})+\oint_C {\phi\over 2} \Bigl(\cot{\phi\over
2}[x-y-2i]\exp[-{2\pi J\sinh \phi\over T\phi}{\bf a}_1(y+i)] \nonum\\
&& +\cot{\phi\over 2}[x-y+2i]\exp[-{2\pi J\sinh \phi\over T\phi} {\bf
a}_1(y-i)]\Bigr){1\over u(y)}{{\rm d}y\over 2\pi i },
\label{eq:neweq}
\enqa 
where the free energy is given by 
\beq f=-T\ln
u(0).
\label{eq:fen2} 
\enq 
The contour $C$ is an arbitrary closed loop
counterclockwise around $0$ where $2nQ, n\ne 0$ and $\pm 2i+2nQ$ should
lie outside of this loop. Furthermore this loop should not contain
zeros of $u(y)$. It is expected that $u(y)$ has no zero in the region
$|\Im y|\le 1$.  We show that these equations can be derived in the
quantum transfer matrix approach.

\section{Quantum transfer matrix and fusion hierarchy models}
The quantum transfer matrix for this model is equivalent to that of the 
diagonal-to-diagonal transfer-matrix of the six-vertex model
which is a staggered or inhomogeneous row-to-row transfer matrix, see below. 
The partition function $Z\equiv {\rm Tr}\exp(-{\cal H}/T)$ is given by
$$Z=\sum_{\{\sigma\}}\prod^N_{j=1}\prod_{i=1}^M 
A(\sigma_{2i+j,j}\sigma_{2i+j+1,j};
\sigma_{2i+j,j+1}\sigma_{2i+j+1,j+1}),$$
$$A(\sigma_1\sigma_2;\sigma_1'\sigma_2')
=\left[
\matrix{
\hfil a&\hfil 0&\hfil 0&\hfil 0\cr
\hfil 0&\hfil c&\hfil b'&\hfil 0\cr
\hfil 0&\hfil b&\hfil c&\hfil 0\cr
\hfil 0&\hfil 0&\hfil 0&\hfil a\cr}\right].$$ 
\beqa
a&\displaystyle =\exp(-{J\Delta \over 2MT})\sinh ({J\over 2MT}),\quad 
b=\exp({-h\over MT}),\nonum\\
b'&\displaystyle =\exp({h \over MT}),\quad 
c=\exp(-{J\Delta \over 2MT})\cosh ({J \over 2MT}).
\enqa 
Then in the case $N=2M\times {\rm integer}$, we have
$$Z={\rm Tr}{\bf T}^N,\quad 
{\bf T}(\sigma_1,\sigma_2,...,\sigma_{2M};
\sigma_1',\sigma_2',...,\sigma_{2M}')$$
$$\equiv A(\sigma_1\sigma_2;\sigma_{2M}'\sigma_1')
A(\sigma_3\sigma_4;\sigma_2'\sigma_3')...
A(\sigma_{2M-1}\sigma_{2M};\sigma_{2M-2}'\sigma_{2M-1}').$$
The eigenvalue problem of this transfer matrix is a special case 
of the
inhomogeneous six-vertex model on the square lattice.

Consider an inhomogeneous six-vertex model with the following 
column dependent Boltzmann weights:
\beqa
&&a_l=\rho_l{\sf h}(v+v_l+\eta)\nonum\\
&&b_l=\rho_l\omega^{-1}{\sf h}(v+v_l-\eta)\nonum\\
&&b_l'=\rho_l\omega{\sf h}(v+v_l-\eta)\nonum\\
&&c_l=\rho_l{\sf h} (2\eta),~~~l=1,...,L.\label{eq:condinhomo}
\enqa
Here $L$ is the number of columns, ${\sf h}(u)$ is $u$, 
$\sin(u)$ or $\sinh(u)$ 
depending on the anisotropy parameter. 
The transfer matrix ${\bf T}(v)$ 
acts in a $2^L$ dimensional space,
\beqa
&&{\bf T}={\rm Tr}{\bf R}_1(\sigma_1,\sigma'_1){\bf R}_2(\sigma_2,\sigma'_2)...
{\bf R}_{L}(\sigma_{L},\sigma'_{L}),\nonum\\
&&{\bf R}_l(++)=\left(\matrix{\hfil a_l&\hfil 0\cr\hfil 0&\hfil b_l\cr}\right),
\quad 
{\bf R}_l(+-)=\left(\matrix{\hfil 0&\hfil 0\cr\hfil c_l&\hfil 0\cr}\right),
\nonum\\
&&{\bf R}_l(-+)=\left(\matrix{\hfil 0&\hfil c_l\cr\hfil 0&\hfil 0\cr}\right),
\quad 
{\bf R}_l(--)=\left(\matrix{\hfil b_l'&\hfil 0\cr\hfil 0&\hfil a_l\cr}\right).
\label{eq:tminh6}
\enqa
The space is divided into
subspaces characterised by the number of down spins $k$. 
Without loss of generality we can put $k\le L/2$. 
In this subspace we can construct Bethe-ansatz wave functions 
with $k$ parameters $u_1,...,u_k$,
\beqa
&&|\Psi\rangle=\sum f(y_1,y_2,...,y_k)\sigma_{y_1}^-\sigma_{y_2}^-...
\sigma_{y_k}^-|0\rangle,\nonum\\
&&f(y_1,y_2,...,y_k)=\sum_PA(P)\prod_{j=1}^k
F(y_j;u_{Pj}),\nonum\\
&&F(y;u)\equiv\omega^y\prod_{l=1}^{y-1}{\sf h}(u+v_l+\eta)\prod_{l=y+1}^L
{\sf h}(u+v_l-\eta),\nonum\\
&&A(P)=\epsilon(P)\sum_{j<l}{\sf  h}(u_{Pj}-u_{Pl}-2\eta).
\enqa
Imposing periodic boundary conditions the Bethe ansatz equations (BAE)
take the form
\beqa
&&{\varphi(u_j+\eta)\over \varphi(u_j-\eta)}=-\omega^{-L}\prod_{m=1}^k
{{\sf h}(u_j-u_m+2\eta)\over  {\sf h}(u_j-u_m-2\eta)},\nonum\\
&&\varphi(v)=\prod_{l=1}^L \rho_l{\sf h}(v+v_l).
\enqa
The corresponding eigenvalue of the transfer matrix is given by
\beqa
&&{\rm T}_1(v)=\omega^{-L+k}\varphi(v-\eta){{\rm  Q}(v+2\eta)\over {\rm  Q}(v)}
 +\omega^{k}\varphi(v+\eta)
{{\rm Q}(v-2\eta)\over {\rm  Q}(v)},\nonum\\
&&{\rm  Q}(v)=\prod_{j=1}^k{\sf h}(v-u_j).
\enqa
In order to solve the diagonal-to-diagonal transfer matrix 
we have to consider an inhomogeneous six-vertex model the Boltzmann 
weights of which are given by
\beqa
&&a_{l}=c_{l}=1,~~b_l=b_l'=0~~~{\rm for~~even~~}l,\nonum\\
&&a_l=\exp\Bigl(-{J\Delta\over 2MT}\Bigr)
\sinh\Bigl({J\over 2MT}\Bigr),~
b_l=\exp({-h\over MT}),\nonum\\
&&b_l'=\exp\Bigl({h\over MT}\Bigr),~~
c_l=\exp\Bigl(-{J\Delta\over 2MT}\Bigr)
\cosh\Bigl({J\over 2MT}\Bigr)
~~~{\rm for~~odd~~}l.
\enqa
The conditions (\ref{eq:condinhomo}) are satisfied if we put
\beq
L=2M,~\omega=\exp\Bigl({h\over MT}\Bigr),~v=0,
~{{\sf h}'( 2\eta)\over {\sf h}'(0)}
={\sinh({J\Delta\over MT})\over \sinh({J\over MT})}, 
\enq
and
\beqa 
&&\rho_l=1/{\sf h}( 2\eta),~ v_l=\eta~~~{\rm for~~even~~}l
\nonum\\
&&\rho_l={\sqrt{bb'}\over {\sf h}( v_l-\eta)},~~
{{\sf h}( v_l+\eta)\over {\sf h}( v_l-\eta)}
={a\over\sqrt{bb'}}=\exp\Bigl(-{J\Delta\over 2MT}\Bigr)
\sinh\Bigl({J\over 2MT}\Bigr)
~~~{\rm for~~odd~~}l.
\enqa  
Putting $\eta+v_1=2\alpha_M$ we have
\beq
\varphi(v)=\Bigl({{\sf h}(v+\eta){\sf h}(v+2\alpha_M-\eta)
\over{\sf h}(2\eta){\sf h}(2\alpha_M-2\eta)}\Bigr)^M.
\label{eq:varphi}\enq
The largest eigenvalue belongs to the $k=M$ sector. The 
Bethe-ansatz equation for $u_j,~j=1,...,M$ are
\beq
{\varphi(u_j+\eta)\over \varphi(u_j-\eta)}
=-e^{-2h/T}\prod_{m=1}^M{{\sf h}(u_j-u_m+2\eta)
\over{\sf h}(u_j-u_m-2\eta)}.\label{eq:dtdBae}
\enq
The corresponding eigenvalue is given by
\beq
{\rm T}_1(v)=e^{-h/T}\varphi(v-\eta){{\rm  Q}(v+2\eta)\over {\rm  Q}(v)}
 +e^{h/T}\varphi(v+\eta){{\rm Q}(v-2\eta)\over {\rm  Q}(v)}.
\label{eq:T(v)}
\enq
Due to the BAE (\ref{eq:dtdBae}), the eigenvalue ${{\rm T}_1(x)}$ is an 
entire function in the complex plane.  The free energy per site is 
given by 
\begin{equation}
f = - T \lim_{M \rightarrow \infty} \ln {\rm T}_1(0).\label{eq:qtmfree}
\end{equation}
The matrix ${{\rm T}_1(v)}$ can be embedded into a more general 
family of matrices provided by the fusion hierarchy \cite{KSS},
\begin{eqnarray}
    {\rm T}_{j}(v)&\equiv& \sum_{l=0}^j e^{-(j-2l)h/T} \varphi(v-(j-2l)\eta)
    \frac{{\rm  Q}(v+(j+1)\eta) {\rm  Q}(v-(j+1)\eta)}{{\rm  Q}(v+(2l-j+1)\eta){\rm  Q}(v+(2l-j-1)\eta)}. 
    \label{eq:Tj(v)}
\end{eqnarray}
The eigenvalues ${\rm T}_{j}(v)$ as functions of $v$ 
are all entire in the complex plane.
It is easily seen that the following functional relations hold \cite{KSS}
\begin{eqnarray}
&&{\rm T}_{j}(v + \eta) {\rm T}_{j}(v -\eta) = \varphi(v+(j+1)\eta)\varphi(v-(j+1)\eta)
 + {\rm T}_{j+1}(v) {\rm T}_{j-1}(v),
\nonumber\\~~
&&{\rm T}_{0}(v) \equiv \varphi(v). 
\label{eq.Tsystem}
\end{eqnarray}

\section{ Derivation of Gaudin-Takahashi equation}
For $\Delta >1$ we put
\beqa
&&{\sf h}(u)= \sin u, \quad \eta=i\tilde{\phi}/2, \quad
\tilde{\phi}=
\cosh^{-1}\bigl({\sinh(J\Delta/2MT)\over 
\sinh(J/2MT)}\bigr), \nonum\\
&&\alpha_M={i\over 2}\tanh^{-1}
\Bigl(\tanh \tilde{\phi}\tanh{J\Delta\over 2MT}\Bigr),\label{eq:5a}
\enqa
In the limit of $M\to \infty$ we have 
\beq
\tilde{\phi}=\phi,~~~~~M\alpha_M=iJ\sinh \phi/(4T). 
\enq
We transform the parameter $v$ to $x\equiv iv/\eta$. 
Then equations (\ref{eq:varphi}) 
and (\ref{eq:Tj(v)}) turn into
\begin{eqnarray}
{\rm  Q}(x) &=& \prod_{j=1}^{M} \sin \frac{\tilde{\phi}}{2}(x - x_j),~~~~
\varphi(x)=\Bigl({\sin{\tilde{\phi}\over 2}(x+ i) \sin{\tilde{\phi}\over 2}
(x-(1-2u_M)i)
\over \sinh \tilde{\phi} \sinh \tilde{\phi}(1-u_M)}\Bigr)^M
,\nonumber \\
u_{M}& =&\alpha_M/\eta, ~~~~~ x_j=iu_j/\eta.  
\end{eqnarray}
\begin{eqnarray}
    {\rm T}_{j}(x)&\equiv& \sum_{l=0}^j e^{-(j-2l)h/T} \varphi(x-(j-2l)i)
    \frac{{\rm  Q}(x+(j+1)i) {\rm  Q}(x-(j+1)i)}{{\rm  Q}(x+(2l-j+1)i){\rm  Q}(x+(2l-j-1)i)}. 
\end{eqnarray}
These functions are all entire in the complex plane.
Now we introduce a modified eigenvalue of ${{\rm T}_j(x)}$
\begin{eqnarray}
\tilde{\rm T}_j(x) &\equiv& {\rm T}_j(x)\Bigl({
\sinh(\tilde{\phi})\sinh \tilde{\phi}(1-u_M) \over 
\sin{\tilde{\phi}\over 2}(x+(j+1)i)
\sin{\tilde{\phi}\over 2}(x-(j+1-2u_M)i)} \Bigr)^M.
\end{eqnarray}
In contrast to the entire function ${{\rm T}_j(x)}$, 
${\tilde{\rm T}_j(x)}$ has poles of order ${M}$ at 
\par\noindent
${x=2nQ+u_Mi\pm(1+j-u_M)i}$. On the other hand, it has constant asymptotics
\begin{equation}
 \tilde{\rm T}_j(\pm  i\infty) =  {\sinh  (j+1)h/T\over \sinh h/T}.
 \end{equation}From 
(\ref{eq.Tsystem}), we can find the following functional relation for 
${\tilde{\rm T}_j(x)}$
\begin{equation}
\tilde{\rm T}_{j}(x+i) \tilde{\rm T}_{j}(x-i) = {\bf  b}_j(x) 
+ \tilde{\rm T}_{j-1}(x)\tilde{\rm T}_{j+1}(x), \label{eq.rel1}
\end{equation}
where we have defined
\begin{eqnarray}
{\bf  b}_j(x)=\Bigl({\sin{\tilde{\phi}\over 2}(x+(j+2u_M)i)
\sin{\tilde{\phi}\over 2}(x-ji)\over 
\sin{\tilde{\phi}\over 2}(x+ji)\sin{\tilde{\phi}\over 2}(x-(j-2u_M)i)}\Bigr)^M.
\end{eqnarray}
Note that ${\tilde{\rm T}_{0}(x)}=1$ and 
${\bf  b}_j(x),~{\tilde{\rm T}_{j}(x)}$ has
poles at ${x=2nQ+u_Mi\pm(j-u_M)i}$ and ~~~~~~~~~~~~~
\par\noindent
${x=2nQ+u_Mi\pm(j+1-u_M)i}$, respectively. 

We define 
\beq
Y_j(x)={\tilde{\rm T}_{j-1}(x)\tilde{\rm T}_{j+1}(x)\over {\bf  b}_j(x)},~~~j=1,2,....
\enq
For these functions the following relations stand
\beqa
&&Y_1(x-i)Y_1(x+i)=1+Y_2(x),\nonum\\
&&Y_j(x+i)Y_j(x-i) = (1+Y_{j-1}(x))(1+Y_{j+1}(x)),~~~j=2,3,..., \nonum\\
&&\lim_{l\to\infty}{\ln Y_l(x)\over l}={2h\over T}.\label{eq:gt3}
\enqa
As $Y_j(x),~j=2,3,...$ has no pole or zero in $-1\le\Im x\le 1$, 
we find
\beq
\ln Y_j(x)={\bf s}*(\ln(1+Y_{j-1})+\ln(1+Y_{j+1})),~~j\ge 2.\label{eq:gt2}
\enq
For $Y_1(x)$ one must be careful that it has poles 
in $-i, (1-2u_M)i$. Using
\beq
\tilde{\rm T}_{2}(x+i) \tilde{\rm T}_{2}(x-i) = {\bf  b}_2(x)(1+Y_2(x)), 
\enq
and $\tilde{\rm T}_{2}(x)$ has no zero or pole at $-1\le\Im x\le 1$, we have
\beq
\ln \tilde{\rm T}_{2}(x)={\bf s}*(\ln {\bf  b}_2(x)+\ln(1+Y_2(x))).
\enq
Using $Y_1(x)=\tilde{\rm T}_2(x)/{\bf  b}_1(x)$ we have
\beq
\ln Y_1(x)=-\ln {\bf  b}_1(x)+{\bf s}*\ln {\bf  b}_2(x)+{\bf s}*\ln(1+Y_2(x)).\label{eq:gt1}
\enq
In the limit of $M\to \infty$, 
the function ${{\bf  b}_j(x)}$ can be 
simplified
\begin{eqnarray}
{\bf  b}_j(x) 
&=& \lim_{M \rightarrow \infty} \exp\Bigl[ M 
\ln \frac{\sin \frac{\tilde{\phi}}{2}(x+(j+2u_M)i) 
\sin \frac{\tilde{\phi}}{2}(x-ji)}
{\sin \frac{\tilde{\phi}}{2}(x+ji) 
\sin \frac{\tilde{\phi}}{2}(x-(j-2u_M)i)}\Bigr] \nonumber \\
&=& \exp \left( -{2\pi J \sinh \phi\over \phi T}{\bf a}_j(x) \right), ~~
{\bf a}_j(x)\equiv{\phi\sinh j\phi/(2\pi)\over \cosh j\phi -\cos(\phi x)},
\label{eq.lambda1}
\end{eqnarray}
which has singularities at ${x=2nQ\pm ji}$. 
In the limit of $M\to \infty$ equations 
(\ref{eq:gt1}),(\ref{eq:gt2}),(\ref{eq:gt3}) are identical to (\ref{eq:GTeq}).
Substituting 
\beq
\ln\tilde{\rm T}_1(x)={\bf s}*\ln[(1+Y_1(x))/{\bf  b}_1(x)]
\enq 
into (\ref{eq:qtmfree}) we have (\ref{eq:fen1}). 
Then the Gaudin-Takahashi equations are derived from the 
quantum transfer matrix method. 
(See also the treatment in
\cite{Klum92,KSS} for related models).

Consider the $M\to\infty$ limit of the functions 
${{\bf  b}_{j}(x), \tilde{\rm T}_{j}(x)}$ as
\begin{equation}
u_j(x) \equiv \lim_{M \rightarrow \infty} \tilde{\rm T}_{j}(x), \ \ 
\end{equation}
Then from (\ref{eq.rel1}) we have the relation
\begin{equation}
u_1(x+i) u_1(x-i) = {\bf  b}_1(x) + u_{2}(x). \label{eq.rel5}
\end{equation}
Note also the asymptotics  ${u_1(\pm i\infty) = 2 \cosh  h/T}$.
We may assume the functions ${u_{1}(x)}$ and ${u_{2}(x)}$ have similar 
singularities at ${x=2nQ\pm 2 i}$ and 
${x=2nQ\pm 3i}$, respectively. If we write (\ref{eq.rel5}) as
\begin{equation}
u_{1}(x+i) = {\bf  b}_1(x)/u_{1}(x-i) + u_{2}(x)/u_{1}(x-i),  \label{eq.rel4}
\end{equation}
the LHS has singularities at ${x = i, -3i}$ in the fundamental region 
(${|\Re x|\le Q}$). The first term of the RHS has singularities at 
${x = i,-i, 3i}$ and the second term at ${x = 3i,-3i -i}$. 
Then following the method in \cite{MT}, we get an integral equation for 
${u_{1}(x)}$,
\begin{eqnarray}
u_1(x) &=& 2 \cosh h/T \nonumber \\
&+& \oint_C \frac{\phi}{2} \Big( \cot \frac{\phi}{2}[x-y-2i] {\bf  b}_1(y+i) + 
\cot \frac{\phi}{2}[x-y+2i] {\bf  b}_1(y-i) \Big) \frac{1}{u_1(y)} 
\frac{{\rm d} y}{2 \pi i}. \nonumber \\ \label{eq.integral}
\end{eqnarray}From 
the explicit expression of ${u_{1}(x)}$ (\ref{eq.lambda1}), we see that 
the integral equation (\ref{eq.integral}) is identical to the one obtained 
in \cite{MT}. 
The free energy is given by 
\begin{equation}
f = - T \ln u_1(0). \label{eq.fenergy}
\end{equation}

\section{Case $\Delta<1$}
In this case we have
\beqa
&&{\sf h}(u)=\sinh u,~~\eta=i\tilde{\theta}/2,~~\tilde{\theta}=
\cos^{-1}\Bigl({\sinh(J\Delta/2MT)\over \sinh(J/2MT)}\Bigr),\nonum\\
&&\alpha_M={i\over 2}\tanh^{-1}\Bigl(
\tan\tilde{\theta}\tanh{J\Delta\over 2MT}\Bigr).
\enqa
In the limit of $M\to \infty$ we have
\beq
\tilde{\theta}=\cos^{-1}\Delta, ~~M\alpha_M=iJ\sin \theta/(4T).
\enq
Putting $x=iv/\eta$ we obtain
\beq
{\rm  Q}(x)=\prod^M_{j=1}\sinh {\tilde{\theta}\over 2}(x-x_j), ~~
\varphi(x)=\Bigl({\sinh{\tilde{\theta}\over 2}(x+ i) \sinh{\tilde{\theta}\over 2}
(x-(1-2u_M)i)
\over \sin \tilde{\theta} \sin \tilde{\theta}(1-u_M)}\Bigr)^M.
\enq
Kuniba, Sakai and Suzuki \cite{KSS} succeeded in deriving the 
Takahashi-Suzuki equations \cite{TS}
for the thermodynamics of the XXZ model at $h=0,~ |\Delta|<1$. 
The functions 
\begin{eqnarray}
\tilde{\rm T}_j(x) &\equiv& {\rm T}_j(x)\Bigl({
\sin(\tilde{\theta})\sin \tilde{\theta}(1-u_M) \over 
\sinh{\tilde{\theta}\over 2}(x+(j+1)i)
\sinh{\tilde{\theta}\over 2}(x-(j+1-2u_M)i)} \Bigr)^M.
\end{eqnarray}
are all periodic with periodicity $2p_0i$. 
We have relations for $\tilde{\rm T}_1(x)$ and $\tilde{\rm T}_2(x)$ 
\begin{equation}
\tilde{\rm T}_{1}(x+i) \tilde{\rm T}_{1}(x-i) = b_1(x) 
+ \tilde{\rm T}_{2}(x), \label{eq.relts}
\end{equation}
with
\begin{eqnarray}
b_1(x)=\Bigl({\sinh{\tilde{\theta}\over 2}(x+(1+2u_M)i)
\sinh{\tilde{\theta}\over 2}(x-i)\over 
\sinh{\tilde{\theta}\over 2}(x+i)
\sinh{\tilde{\theta}\over 2}(x-(1-2u_M)i)}\Bigr)^M.
\end{eqnarray}
$\tilde{\rm T}_1(x)$ satisfies
\begin{equation}
 \tilde{\rm T}_1(\pm  \infty) =  {2\cosh h/T}.
 \end{equation}
By these two equations we can determine $\tilde{\rm T}_1(x)$ 
in the limit of $M\to \infty$. In this limit $b_1(x)$ is 
\beq
b_1(x)=\exp \left( -{2\pi J \sin \theta\over \theta T}{ a}_1(x) \right), ~~
{ a}_1(x)\equiv{\theta\sin \theta/(2\pi)\over \cosh (\theta x) -\cos\theta}.
\enq
We can assume that $\tilde{\rm T}_1(x)$ is expanded as follows
\beq
\tilde{\rm T}_1(x)=2\cosh({h\over T})+\sum_{j=1}^{\infty}
\sum_n{c_j\over (x-2np_0i-2i)^j}
+\sum_{j=1}^{\infty}
\sum_n{\overline{c_j}\over(x-2np_0i+2i)^j}.\label{eq:cj}
\enq
Consider the contour integral around $x=i$ giving the coefficients $c_j$ 
\beq
c_j=\oint {(x-i)^{j-1}b_1(x)\over \tilde{\rm T}_1(x-i)}
{{\rm d}x\over 2\pi i}
=\oint {y^{j-1}b_1(y+i)\over \tilde{\rm T}_1(y)}{{\rm d}y\over 2\pi i}.
\enq
The first sum of the r.h.s. of (\ref{eq:cj}) is
\beqa
&&\sum_{j=1}^\infty 
\oint\sum_n{b_1(y+i)\over (x-2np_0i-2i)^j}
{y^{j-1}\over \tilde{\rm T}_1(y)}{{\rm d}y\over 2\pi i}
=\oint \sum_n
{b_1(y+i)\over x-y-2np_0i-2i}
{1\over \tilde{\rm T}_1(y)}{{\rm d}y \over 2\pi i}\nonum\\
&&=\oint {\theta\over 2}\coth{\theta\over 2}(x-y-2i)
\exp[-{2\pi J\sin \theta\over T\theta}{ a}_1(y+i)]
{1\over \tilde{\rm T}_1(y)}{{\rm d}y\over 2\pi i}. 
\enqa
The second sum is calculated in a similar way. Thus we find
\beqa
&&u(x)=2\cosh({h\over T})+\oint_C {\theta\over 2}
\Bigl(\coth{\theta\over 2}[x-y-2i]\exp[-{2\pi J\sin \theta\over T\theta}
{ a}_1(y+i)]
\nonum\\
&&
+\coth{\theta\over 2}[x-y+2i]\exp[-{2\pi J\sin \theta\over T\theta}
{ a}_1(y-i)]\Bigr){1\over u(y)}{{\rm d}y\over 2\pi i },\label{eq:neweq2}
\enqa
and the free energy is given by
\beq
f=-T\ln u(0).
\enq
Apparently these equations are analytical continuations of (\ref{eq:neweq}) 
and (\ref{eq:fen2}) if we replace $\phi$ by $i\theta$. Then equation 
(\ref{eq:neweq}) treats the thermodynamics in a unified way. 

This research is supported in part by Grants-in-Aid for the Scientific
Research (B) No. 11440103 from the Ministry of Education, Science and
Culture, Japan and Japan-German cooperation program of scientific
research "Theory of low-dimensional quantum spin systems and
correlated electron systems" by JSPS and DFG.

\end{document}